\documentclass[final]{aipproc}
\layoutstyle{8x11double}
\newcommand\doingARLO[2][]{%
  \ifx\mmref\undefined #1\else #2\fi
}

\begin{document}
\title[The ROTSE-IIIa Telescope System]{The ROTSE-IIIa Telescope System}

\author{D. Smith}{
	address={2477 Randall Laboratory, University of Michigan, 500
	E. Univeristy Ave., Ann Arbor, MI, 48109, USA},
	email={donaldas@umich.edu}
}
\author{C. Akerlof}{
	address={2477 Randall Laboratory, University of Michigan, 500
	E. Univeristy Ave., Ann Arbor, MI, 48109, USA},
	email={akerlof@umich.edu}
}
\author{M. C. B. Ashley}{
	address={School of Physics, University of New South Wales, Sydney 2052,
	Australia},
	email={mcba@phys.unsw.edu.au}
}
\author{D. Casperson}{
	address={Los Alamos National Laboratories, Los Alamos, NM, 87545, USA},
	email={dcasperson@lanl.gov}
}
\author{G. Gisler}{
	address={Los Alamos National Laboratories, Los Alamos, NM, 87545, USA},
	email={ggisler@lanl.gov}
}
\author{R. Kehoe}{
	address={2477 Randall Laboratory, University of Michigan, 500
	E. Univeristy Ave., Ann Arbor, MI, 48109, USA},
	email={rlkehoe@umich.edu}
}
\author{S. Marshall}{
	address={Lawrence Livermore National Laboratory, 7000 East Avenue,
	Livermore, CA, 94550, USA},
	email={stuart@igpp.llnl.gov}
}
\author{K. McGowan}{
	address={Los Alamos National Laboratories, Los Alamos, NM, 87545, USA},
	email={mcgowan@aslan.lanl.gov}
}
\author{T. McKay}{
	address={2477 Randall Laboratory, University of Michigan, 500
	E. Univeristy Ave., Ann Arbor, MI, 48109, USA},
	email={tamckay@umich.edu}
}
\author{M. A. Phillips}{
	address={School of Physics, University of New South Wales, Sydney 2052,
	Australia},
	email={a.phillips@unsw.edu.au}
}
\author{E. Rykoff}{
	address={2477 Randall Laboratory, University of Michigan, 500
	E. Univeristy Ave., Ann Arbor, MI, 48109, USA},
	email={erykoff@umich.edu}
}
\author{W. T. Vestrand}{
	address={Los Alamos National Laboratories, Los Alamos, NM, 87545, USA},
	email={vestrand@lanl.gov}
}
\author{P. Wozniak}{
	address={Los Alamos National Laboratories, Los Alamos, NM, 87545, USA},
	email={wozniak@algol.lanl.gov}
}
\author{J. Wren}{
	address={Los Alamos National Laboratories, Los Alamos, NM, 87545, USA},
	email={jwren@nis.lanl.gov}
}

\copyrightyear{2001}

\begin{abstract}

We report on the current operating status of the ROTSE-IIIa telescope,
currently undergoing testing at Los Alamos National Laboratories in New Mexico.
It will be shipped to Siding Spring Observatory, Australia, in first quarter
2002.  ROTSE-IIIa has been in automated observing mode since early October,
2001, after completing several weeks of calibration and check-out observations.
Calibrated lists of objects in ROTSE-IIIa sky patrol data are produced
routinely in an automated pipeline, and we are currently automating analysis
procedures to compile these lists, eliminate false detections, and
automatically identify transient and variable objects.  The manual application
of these procedures has already led to the detection of a nova that rose over
six magnitudes in two days to a maximum detected brightness of $m_{\rm
R}\sim13.9$ and then faded two magnitudes in two weeks.  We also readily
identify variable stars, includings those suspected to be variables from the
Sloan Digital Sky Survey.  We report on our system to allow public monitoring
of the telescope operational status in real time over the WWW.

\end{abstract}

\maketitle

\section{Telescope Status}

ROTSE-IIIa is a 0.45-m robotic reflecting telescope which is managed by a
fully-automated system of interacting daemons within a Linux environment.  The
telescope has an f-ratio of 1.9, yielding a field of view (FOV) of
$1.85\times1.85$ degrees.  The control system is connected via a TCP/IP socket
to the Gamma-Ray Burst Coordinate Network (GCN), and a flexible scheduler
daemon plans observation sequences including sky patrols, targeted monitoring
programs, and fast ($<10$~s) responses to GRB alerts.  Upon receipt of a GRB
alert over the GCN, we begin a program of 10 5-s, 10 20-s, and 80 60-s
exposures (with a read-out time of 7~s between images).  We also automatically
schedule blocks of 30 60-s follow-up exposures.  These blocks are spaced at
ever-increasing intervals.  The system began its testing run on October 11,
2001, at the ROTSE-I site at Los Alamos National Laboratory.  ROTSE-IIIa will
be shipped to Siding Springs Observatory, Australia, upon completion of the
testing phase.

\begin{figure}
\caption{Number of objects per magnitude bin for a random ROTSE-IIIa field
at three exposure times.  Magnitudes were derived by calibrating to the USNO
R-band and then applying a constant correction factor of 0.3~mag to convert to
an approximate V-band.}
\includegraphics[height=.25\textheight]{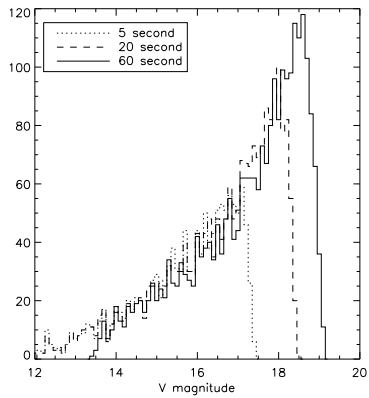}
\label{fig:bsen}
\end{figure}

ROTSE-IIIa currently uses unfiltered CCDs, although we have included a slot in
the mechanical design that will allow for the insertion of a filter.  We
currently calibrate ROTSE magnitudes against the USNO R-band magnitudes, and we
include a constant offset of 0.3 magnitudes to convert these numbers to a
V-band magnitude.  This necessarily introduces unknown systematic errors for
objects with atypical spectra.  As shown in Figure~\ref{fig:bsen}, ROTSE-IIIa
can reach 17th magnitude in a 5-s exposure, 17.5 in 20-s, and 18.5 in a 60-s
exposure.  Longer exposures are not practical due to saturation of the sky.
Multiple images can be co-added to reach $\sim19$th magnitude.

Figure~\ref{fig:phom} shows the dispersion of calibrated stellar magnitudes
over time (corrected to V band).  Plotted are the standard deviations vs. the
mean magnitude of $\sim3000$ objects detected in at least 10 of 20 observations
of a single field.  The intensity of bright stars varies by less than 1\%,
while the RMS deviation rises to $\sim30$\% for the dimmest stars ROTSE-IIIa
can detect.

To measure the astrometric accuracy of our analysis, we compare the relative
positions of objects found in both the USNO~A2.0 catalog as well as in the
abovementioned 20 ROTSE-IIIa observations.  The transformation of the CCD
coordinates to celestial coordinates is achieved through a third-oder
polynomial warp.  Most bright stars ($m_{\rm V}<14$) can be localized to better
than 0.3 arcsec, or one-tenth of a pixel, and the bulk of the faintest objects
stay within two-thirds of a pixel.

\section{Data Analysis Pipeline}

We are implementing software to automatically analyze all images recorded with
the ROTSE-III telescopes.  The full analysis pipeline for each telescope will
be installed on a computer at each site, but during testing we are copying most
of the data to the University of Michigan for easier control.  All images are
automatically dark- and flat-field corrected as soon as they are recorded.
Immediately after each corrected image file is written, it is processed into an
object list with the SExtractor package~\cite{sextr}, using an aperture 5
pixels in diameter, to identify all source candidates within the FOV.  The list
of these candidate objects is compared against the USNO~A2.0 catalog using a
triangle-matching routine to compile a calibrated list of R-band magnitudes and
celestial locations for these sources.  Calibrated object lists also allow
diagnostic parameters to be measured, such as the astrometric accuracy and the
focus quality.  These steps have been implemented on an on-site computer and
the system produces a calibrated object list for each image within 45~s from 
the closing of the shutter.

\begin{figure}
\caption{Photometric scatter in 20 60-s observations of 3,000 objects over a
two-week interval as a function of V-band magnitude. The images have been
corrected using a relative photometry algorithm.  Root mean square deviations
vary from $\sim1$\% at 14th magnitude to $\sim10-30$\% at the magnitude limit.}
\includegraphics[height=.25\textheight]{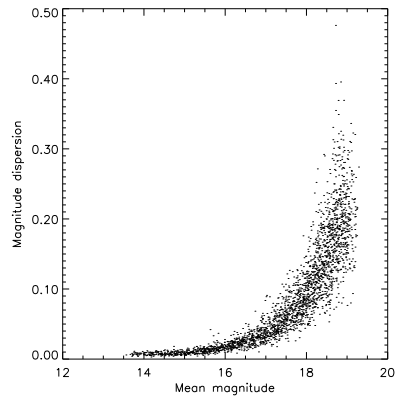}
\label{fig:phom}
\end{figure}

Once more than one calibrated list for a given field is available, these lists
are compiled into a ``match structure''; a data structure that enables us to
filter out objects that do not appear at the same location in sequential
observations.  The telescope aspect is shifted through a small, random vector
between observations.  This enables the elimination of hot CCD pixels and
cosmic ray events.  We then apply a relative photometry algorithm to stabilize
the magnitude estimates and calibrate the systematic errors.  This reduces the
scatter in light curves for stable bright objects to $<1$\%
(Fig.~\ref{fig:phom}) and allows us to reliably identify variable sources.  We
can also flag known distractions such as asteroids and ``masked'' stars (stars
that appear in ROTSE images but are too close to bright stars for sensitive
cataloged surveys to resolve).  Once this procedure is automated, we anticipate
that the system will be able to report an arcsecond position for a bright
($m_{\rm R}>17$), variable object within five minutes of the receipt of a burst
trigger via a GCN alert.

\section{Discoveries}

\begin{figure}
\caption{Folded light curve in calibrated R-band for an identified variable star in
one of the ROTSE-IIIa sky patrol fields.  An estimate of a systematic error of
4\% has been added in quadrature to the statistical errors.  The mean magnitude
of 14.75 has been subtracted.}
\includegraphics[height=.22\textheight]{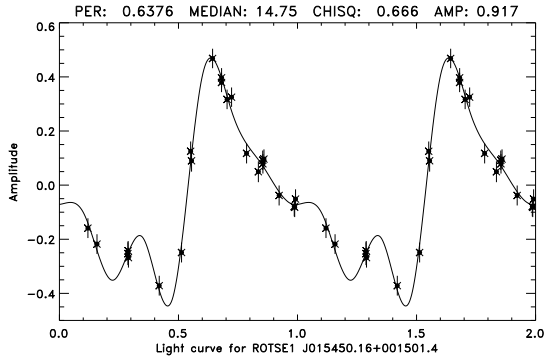}
\label{fig:rrly}
\end{figure}

Calibrated match structures can easily be searched for previously unknown
objects.  For our first testing of the procedures involved, we constructed
these structures for 20 epochs over two weeks for $\sim500$ sq. deg. of sky
patrols along the Equatorial strips of the Sloan Digital Sky Survey Early Data
Release.  Variable stars can be identified through application of the
techniques developed for ROTSE-I~\cite{timvar}.  In Figure~\ref{fig:rrly}, we
show the folded light curve for a typical RR Lyra star in a ROTSE-IIIa field.
The best-fit period of 0.64~d was found using a cubic spline
method~\cite{phasefit}.  This method provides a best-fit period and error
estimate for a variable star light curve, as well as a spline-interpolated
approximation for the source light curve.

\begin{figure}
\caption{Light curve in the calibrated R-band for a transient nova discovered 
in the first two weeks of automated sky patrols by ROTSE-IIIa.  Arrows
indicate the limiting magnitudes of two images taken 1.8 d before the first
detection at $m_{\rm R}=14.00\pm0.04$ on Oct~13.291, 2001.  The source was
easily detectable for the following two weeks, until proximity to the waxing
moon interfered with continued monitoring.}
\includegraphics[height=.35\textheight]{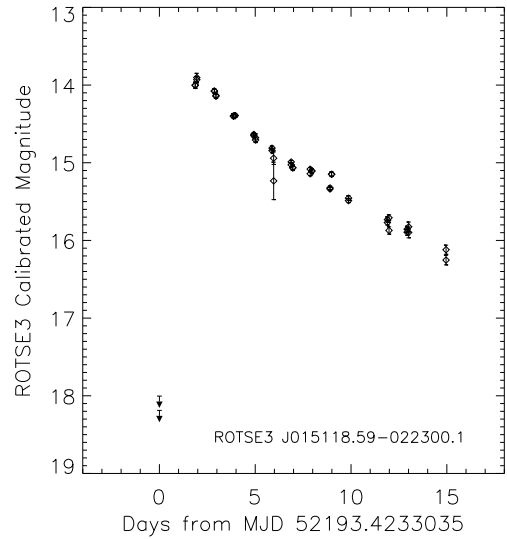}
\label{fig:tran}
\end{figure}

Non-periodic variables can also be identified.  Figure~\ref{fig:tran} shows the
light curve for a new transient that was not detected on Oct 11.424, 2001 (with
a limiting magnitude of 18.2), but was easily visible at a magnitude of 14.0 on
Oct 13.291, 45~h later.  Clouds prevented observations on the night of Oct 12.
The transient then faded two magnitudes over the subsequent two weeks until
proximity to the waxing moon interfered with further monitoring.  The best-fit
power law index for the fading light curve is $\sim0.9$, but the shape of the
curve is clearly not consistent with a single power law.  With a mean magnitude
of 15 and an RMS deviation around that mean of 0.66, this source stood out
starkly against the typical behavior of other stars (Fig.~\ref{fig:phom}).  The
ease with which this source was detected is a positive indication of the
potential for using ROTSE-IIIa to identify and track new transients.

\section{Realtime WWW Monitoring}

The telescope operating system writes all values of its status variables to a
file once every minute.  These variables include the current pointing
direction, the name of the most recent image, whether the enclosure roof is
open or closed, the current velocity of the mount along its two axes, and
whether or not any alarms are active.  This file is copied back to the
University of Michigan by a cron job and parsed into an HTML file that can be
viewed in real time over the WWW at \url{http://www.rotse.net}.  Also included
in this display are a thumbnail of the most recent image and graphs of the
diagnostic parameters derived from the calibrated object lists.  This interface
allows interested parties to monitor burst response and telescope status in
near-real-time.

\section{Conclusions}

ROTSE-IIIa is working well in its first testing phase at Los Alamos.  Operation
and image processing has been automated, and image analysis is being automated
at the time of this writing.  When full automation is implemented, we will be
able to report an arcsecond position for a bright ($m_{\rm R}>17$), rapidly
variable ($\Delta m > 10$\% within the first minute) object within five minutes
of receipt of a burst trigger.  Limiting magnitudes under good conditions in
60-s exposures can approach 19th magnitude, and we anticipate gaining further
sensitivity at the (darker) permanent site in Australia.

\begin{theacknowledgments}
D. Smith is supported by NSF fellowship 00-136.  Work performed at LANL is
supported by NASA SR\&T through Department of Energy (DOE) contract
W-7405-ENG-36 and through internal LDRD funding. Work performed at the
University of Michigan is supported by NASA under SR\&T grant NAG5-5101, the
NSF under grants AST 97-03282 and AST 99-70818, the Research Corporation, the
University of Michigan, and the Planetary Society. Work performed at LLNL is
supported by NASA SR\&T through DOE contract W-7405-ENG-48.

\end{theacknowledgments}

\doingARLO[\bibliographystyle{aipproc}]
          {\ifthenelse{\equal{\AIPcitestyleselect}{num}}
             {\bibliographystyle{arlonum}}
             {\bibliographystyle{arlobib}}
          }
\bibliography{woodshole}

\hyphenation{Post-Script Sprin-ger}
\begin{thebibliography}{3}
\expandafter\ifx\csname natexlab\endcsname\relax\def\natexlab#1{#1}\fi
\providecommand{\enquote}[1]{``#1''}
\expandafter\ifx\csname url\endcsname\relax
  \def\url#1{\texttt{#1}}\fi
\expandafter\ifx\csname urlprefix\endcsname\relax\def\urlprefix{URL }\fi

\bibitem[{Bertin} and {Arnouts}(1996)]{sextr}
{Bertin}, E., and {Arnouts}, S., \emph{A\&AS}, \textbf{117}, 393--404 (1996).

\bibitem[{Akerlof} et~al.(2000)]{timvar}
{Akerlof}, C., et~al., \emph{AJ}, \textbf{119}, 1901--1913 (2000).

\bibitem[{Akerlof} et~al.(1994)]{phasefit}
{Akerlof}, C., et~al., \emph{ApJ}, \textbf{436}, 787--794 (1994).

\end{thebibliography}

\end{document}